\documentclass[a4paper,12pt]{article}

\usepackage{graphics, epsfig}

\begin{document}

\author{P. A. Hogan\thanks{E-mail : peter.hogan@ucd.ie} and D. M. Walsh\thanks{E-mail
: darraghmw@hotmail.com}\\ \small Mathematical Physics
Department,\\ \small National University of Ireland Dublin,
Belfield, Dublin 4, Ireland}

\title{Collision of High Frequency Plane Gravitational and Electromagnetic Waves}
\date{}
\maketitle

\begin{abstract}
We study the head--on collision of linearly polarized, high
frequency plane gravitational waves and their electromagnetic
counterparts in the Einstein--Maxwell theory. The post--collision
space--times are obtained by solving the vacuum Einstein and
Einstein--Maxwell field equations in the geometrical optics
approximation. The head--on collisions of all possible pairs of
these systems of waves is described and the results are then
generalized to non--linearly polarized waves which exhibit the
maximum two degrees of freedom of polarization.
\end{abstract}
\thispagestyle{empty}
\newpage

\section{Introduction}\indent
The primary purpose of this work is to determine the vacuum
gravitational field produced by the head--on collision of high
frequency plane gravitational waves. Although a large number of
collision space--times are known (see for example \cite{G}), this
is a collision scenario which seems so far not to have received
attention. High frequency gravitational waves are produced in the
final stages of a binary neutron star merger, for example, when
the orbital period of the stars is of the order of 100Hz. Such
radiation will ultimately be detected as high frequency plane
waves by a new generation of distant earth--bound detectors. A
qualitative analytical description of the process of high
frequency radiation, in--fall and merger, followed by low
frequency ring--down to the formation of a black hole by an
isolated gravitating system is given in \cite{FH1}. With estimates
of between three and one hundred such cataclysmic events per year
detectable within a distance of hundreds of Mpc \cite{NPS}
\cite{P} it is of some interest to explore the characteristics of
the gravitational field produced by colliding high frequency
gravitational waves

\setcounter{equation}{0}
\section{High Frequency Plane Waves}\indent
To describe high frequency plane gravitational waves we begin with
the space--time model of the gravitational field of linearly
polarized plane gravitational waves propagating through a vacuum.
This space--time has a line--element which can be put in Rosen
\cite{R} form
\begin{equation}\label{1}
ds^2=-F^2dx^2-G^2dy^2+2\,du\,dv\ ,
\end{equation}
where $F, G$ are functions of $u$ only. Einstein's vacuum field
equations are equivalent to \begin{equation}\label{2} \ddot
F-h\,F=0\ ,\qquad \ddot G+h\,G=0\ .\end{equation} Here and
throughout dots indicate differentiation with respect to $u$ and
$h(u)$ is the freely specifiable profile of the waves. If in place
of $F(u),\ G(u)$ we use $B(u),\ w(u)$ given by $F=B\,{\rm e}^w,\
G=B\,{\rm e}^{-w}$ then (\ref{1}) becomes
\begin{equation}\label{3}
ds^2=-B^2({\rm e}^{2\,w}dx^2+{\rm e}^{-2\,w}dy^2)+2\,du\,dv\ .
\end{equation}
Calculation of the Ricci tensor with the metric tensor given via
this line--element results in
\begin{equation}\label{4}
R_{ij}=2\,\left (\frac{\ddot B}{B}+\dot w^2\right
)\,u_{,i}\,u_{,j}\ \ , \end{equation} where the comma denotes
partial differentiation with respect to the coordinates $x^i=(x,
y, u, v)$. On account of (\ref{2}) we have $R_{ij}=0$ and thus
\begin{equation}\label{5}
\ddot B+\dot w^2B=0\ .
\end{equation}In Newman--Penrose notation the only non--vanishing
component of the Riemann curvature tensor is
\begin{equation}\label{6}
\Psi _4=-\ddot w-2\,\dot w\,\frac{\dot B}{B}=-h(u)\ ,
\end{equation} with the last equality here coming from (\ref{2}).
This is a Petrov type N curvature tensor (the type associated with
pure gravitational radiation) having the vector field $\partial
/\partial v$ as four--fold degenerate principal null direction. It
is readily seen from (\ref{1}) or (\ref{3}) that the integral
curves of $\partial /\partial v$ are shear--free, expansion--free
and twist--free null geodesics that generate the null
hypersurfaces $u={\rm constant}$. These null hypersurfaces are the
histories of the wave fronts. They are null hyper\emph{planes} and
the property of the field (\ref{6}) that $\Psi _4={\rm constant}$
when $u={\rm constant}$ distinguishes them as the histories of
\emph{plane} gravitational waves.

The specialization of these plane waves to high frequency plane
waves has been carried out in the following elegant fashion by
Burnett \cite{B}: First replace (\ref{3}) by a family of
line--elements parametrized by a real parameter $\lambda \geq 0$.
Thus we write
\begin{equation}\label{7}
ds^2=-B_{\lambda}^2(u)({\rm e}^{2\,w_{\lambda}(u)}dx^2+{\rm
e}^{-2\,w_{\lambda}(u)}dy^2)+2\,du\,dv\ ,
\end{equation} with \begin{equation}\label{8}
w_{\lambda}(u)=\lambda\,\alpha _0(u)\,\sin \frac{u}{\lambda}\ ,
\end{equation}
and
\begin{equation}\label{9}
\ddot B_{\lambda}+\dot w_{\lambda}^2B_\lambda =0\ .
\end{equation}
Here $\alpha _0(u)$ is an arbitrary (integrable) function of $u$
for some range of $u$ that includes $u=0$. We shall be interested
in these plane waves for small values of $\lambda >0$, for which
$w_\lambda (u)=O(\lambda )$, and also in the limit
$\lambda\rightarrow 0$. We think of $\lambda$ as representing the
wave--length of the waves. We assume that
$\lim_{\lambda\rightarrow 0}B_\lambda (u)=B_0(u)$ exists and is
uniform on the range of $u$. Writing (\ref{9}) as an integral
equation using an appropriate Green's function, taking the limit
$\lambda\rightarrow 0$ and using the Riemann--Lebesgue theorem,
yields an integral equation for $B_0(u)$ from which one can show
that $B_0(u)$ satisfies the differential equation
\begin{equation}\label{10}
\ddot B_0+\frac{1}{2}\,\alpha _0^2B_0=0\ .
\end{equation}
The procedure described here leading to (\ref{10}) is given in
detail in \cite{FH}. For $\lambda >0$ and small we can write the
Isaacson--type expansion \cite{I}
\begin{equation}\label{11}
B_\lambda (u)=B_0(u)+\lambda\,B_1(u, \lambda )+\lambda ^2B_2(u,
\lambda )+\dots\ \ ,
\end{equation}
with $B_n(u, \lambda )=O(\lambda ^0)\ ,\ \dot B_n=O(\lambda
^{-1})\ ,\ \ddot B_n=O(\lambda ^{-2})$ for $n=1, 2, 3, \dots$ .
For future use we shall require that $B_\lambda (0)=1$ for all
$\lambda\geq 0$. It thus follows from (\ref{8}) and (\ref{9}) that
$B_1=0$ and so we have, for small $\lambda$,
\begin{equation}\label{12}
B_\lambda (u)=B_0(u)+O(\lambda ^2)\ .
\end{equation} Substituting (\ref{8}) and (\ref{12}) into the
line--element (\ref{7}) we see that for small $\lambda$ the
line--element splits into a `background' and a small perturbation
as
\begin{equation}\label{13}
ds^2=ds_0^2-2\,\lambda\,B_0^2(u)\,\alpha
_0(u)\,\sin\frac{u}{\lambda}\,(dx^2-dy^2)+O(\lambda ^2)\ ,
\end{equation}
where $ds_0^2$ is the line--element of the `background'
space--time given by \begin{equation}\label{14}
ds_0^2=-B_0^2(u)\,(dx^2+dy^2)+2\,du\,dv\ .
\end{equation}
The space--time with line--element (\ref{7}) is a vacuum
space--time to all orders in $\lambda$. The background space--time
with line--element (\ref{14}) is \emph{not} a vacuum space--time.
The Ricci tensor calculated with the metric tensor given via
(\ref{14}) has components
\begin{equation}\label{15}
R_{ij}^{(0)}=-\alpha _0^2(u)\,u_{,i}\,u_{,j}\ \ .
\end{equation}
The Riemann tensor of the space--time with line--element
(\ref{13}) has one non--vanishing Newman--Penrose component (given
by (\ref{6}) with $w,\ B$ replaced by $w_\lambda ,\ B_\lambda$)
which for small $\lambda$ reads
\begin{equation}\label{16}
\Psi _4=\frac{1}{\lambda}\,\alpha
_0(u)\,\sin\frac{u}{\lambda}+O(\lambda ^0)\ .
\end{equation}The key elements here which one associates with the
geometrical optics approximation are the strong type N
gravitational field of the waves given by (\ref{16}) for small
$\lambda$ (this field then determines the `rays' associated with
the waves, which are the integral curves of the degenerate
principal null direction $\partial /\partial v$), and the
classical form of the line--element (\ref{13}) as a small
perturbation of a background whose Ricci tensor is proportional to
the `square' of the (covariant) propagation direction in
space--time of the histories of the wave fronts. In section 3 we
will derive the approximate vacuum space--time model of the
gravitational field which arises following the head--on collision
of two families of these high frequency plane waves.

As a final preliminary we note that the Einstein--Maxwell vacuum
field equations with linearly polarized plane electromagnetic
waves as source are satisfied by a metric tensor given by the
conformally flat line--element
\begin{equation}\label{17}
ds^2=-B^2(u)\,(dx^2+dy^2)+2\,du\,dv\ ,
\end{equation}
and a type N Maxwell field with one non--vanishing Newman--Penrose
component $\phi _2(u)$ which is real--valued and is given by
\begin{equation}\label{18}
\ddot B+\phi _2^2B=0\ .
\end{equation}
If the electromagnetic waves are not linearly polarized and thus
have the maximum two degrees of freedom of polarization then $\phi
_2(u)$ is a complex--valued function of the real variable $u$ and
(\ref{18}) is replaced by
\begin{equation}\label{19}
\ddot B+|\phi _2|^2B=0\ .
\end{equation}
Thus the high frequency case here, analogous to the high frequency
gravitational waves described above, is obtained by replacing
$B(u)$ by $B_\lambda (u)$ for $\lambda\geq 0$ and taking
\begin{equation}\label{20}
\phi _2(u)=a_0(u)\,\cos\frac{u}{\lambda}=O(\lambda ^0)\ ,
\end{equation}
with $a _0$ an arbitrary real--valued function of $u$ in the
linearly polarized case and a complex--valued function of $u$ in
the general case. Also for small $\lambda$ the line--element
(\ref{17}) takes the form \begin{equation}\label{21}
ds^2=ds_0^2+O(\lambda ^2)\ ,\qquad
ds_0^2=-B_0^2(u)(dx^2+dy^2)+2\,du\,dv\ ,\end{equation}with
$B_0(u)$ satisfying \begin{equation}\label{22a} \ddot
B_0+\frac{1}{2}a_0^2B_0=0\ .
\end{equation} It is interesting
to consider the head--on collision problem for these waves. This
is discussed in section 4 where we also consider the case in which
these waves share their wave fronts with the high frequency
gravitational waves.

The collisions of the linearly polarized gravitational and
electromagnetic plane waves described in sections 3 and 4 can be
generalized to allow the maximum two degrees of freedom of
polarization in each system of waves. This generalization is given
in section 5.

\setcounter{equation}{0}
\section{Collision of Gravitational Waves}\indent
The line--element of the space--time containing the histories of
the wave fronts of the incoming high frequency plane gravitational
waves, as well as modeling their gravitational fields and the
gravitational field after the waves collide, has the
Rosen--Szekeres form
\begin{equation}\label{23}
ds^2=-{\rm e}^{-U}({\rm e}^Vdx^2+{\rm e}^{-V}dy^2)+2\,{\rm
e}^{-M}du\,dv\ ,
\end{equation}
with $U, V, M$ functions of $u, v$ in general. To set up the
head--on collision of linearly polarized high frequency plane
gravitational waves we take the line--element (\ref{23}) in the
region of space--time $u\geq 0, v\leq 0$ to coincide with
(\ref{7}) above. Thus in this region \begin{equation}\label{24}
U=-2\,\log B_\lambda (u)\ ,\ V=2\,w_\lambda (u)=2\,\lambda\,\alpha
_0(u)\,\sin\frac{u}{\lambda}\ ,\ M=0\ .
\end{equation}
For the region $u\leq 0,\ v\geq 0$ we have similar waves traveling
in the opposite direction by taking
\begin{equation}\label{25}
U=-2\,\log D_\lambda (v)\ ,\ V=2\,k_\lambda (v)=2\,\lambda\,\beta
_0(v)\,\sin\frac{v}{\lambda}\ ,\ M=0\ .
\end{equation}
In the region $u\leq 0,\ v\leq 0$ we have flat space--time with
$U=V=M=0$. We thus have these families of plane waves propagating
into a region free of any gravitational field and then engaging in
a head--on collision. To find the line--element of the space--time
in the region $u>0,\ v>0$ after the collision we solve Einstein's
vacuum field equations for $U(u, v),\ V(u, v),\ M(u, v)$ subject
to the boundary conditions: when $v=0$ and $u\geq 0$ the functions
$U, V, M$ are given by (\ref{24}) and when $u=0$ and $v\geq 0$ the
functions $U, V, M$ are given by (\ref{25}). We will be content to
satisfy the vacuum field equations approximately for small
$\lambda$ in the form
\begin{equation}\label{26}
R_{ij}=O(\lambda )\ .
\end{equation}
Thus we look for $U, V, M$ in $u>0,\ v>0$ satisfying the boundary
conditions above and the equations (see \cite{G}, p.39)
\begin{equation}\label{27}
2\,U_{uu}-U_u^2-V_u^2+2\,U_u\,M_u=O(\lambda )\ ,
\end{equation}
\begin{equation}\label{28}
2\,U_{vv}-U_v^2-V_v^2+2\,U_v\,M_v=O(\lambda )\ ,
\end{equation}
\begin{equation}\label{29}
2\,V_{uv}-U_u\,V_v-U_v\,V_u=O(\lambda )\ ,
\end{equation}
\begin{equation}\label{30}
2\,M_{uv}+U_u\,U_v-V_u\,V_v=O(\lambda )\ ,
\end{equation}
and \begin{equation}\label{31} U_{uv}-U_u\,U_v=O(\lambda )\ .
\end{equation}
Here the subscripts denote partial derivatives. These are the
exact vacuum field equations if the $O(\lambda )$--terms are
replaced by zeros. In fact (\ref{31}) with zero on the right hand
side is well known to yield $U$ in the form
\begin{equation}\label{32}
U=-\log (f(u)+g(v))\ ,
\end{equation}
for some functions $f(u),\ g(v)$. To ensure a smooth transition to
Minkowskian space--time in $u<0,\ v<0$ on the boundaries $u=0,\
v\leq 0$ and $v=0,\ u\leq 0$ of this region, we must have the
functions $B_\lambda (u),\ D_\lambda (v)$ in (\ref{24}) and
(\ref{25}) satisfy
\begin{equation}\label{33}
B_\lambda (0)=D_\lambda (0)=1\ ,
\end{equation}
for $\lambda\geq 0$. Indeed to ensure that no impulsive
light--like signals exist on the boundaries $u=0,\ -\infty
<v<+\infty$ and $v=0,\ -\infty <u<+\infty$ we take
\begin{equation}\label{34}
\dot B_\lambda (0)=\dot D_\lambda (0)=\alpha _0(0)=\beta _0(0)=0\
.
\end{equation}
We also note that $B_\lambda (u)$ satisfies (\ref{9}) and
(\ref{12}) with $B_0(u)$ satisfying (\ref{10}), while $D_\lambda
(v)$ satisfies corresponding equations
\begin{equation}\label{35}
\ddot D_\lambda +\dot k_\lambda ^2D_\lambda =0\ ,\qquad \ddot
D_0+\frac{1}{2}\beta _0^2D_0=0\ ,
\end{equation}
and
\begin{equation}\label{36}
D_\lambda (v)=D_0(v)+O(\lambda ^2)\ ,
\end{equation}
with a dot on a function of one variable denoting differentiation
with respect to that variable.

Using the boundary conditions on $U$ along with (\ref{32}) and
(\ref{33}) we see that in the region $u>0,\ v>0$ we have
\begin{equation}\label{37}
U=-\log\{B_\lambda ^2(u)+D_\lambda ^2(v)-1\}\ .
\end{equation}
This holds for $\lambda\geq 0$. On account of (\ref{12}) and
(\ref{36}) we can write this as
\begin{equation}\label{38}
U=-\log\{B_0^2(u)+D_0^2(v)-1\}+O(\lambda ^2)\ ,
\end{equation}
which is sufficiently accurate for our purposes, although the
exact expression (\ref{37}) is also very useful to have. With $U$
given by (\ref{38}) we can solve (\ref{29}) for $V(u, v)$ to
arrive at
\begin{equation}\label{39}
V=\frac{2\,\lambda}{\sqrt{B_0(u)^2+D_0(v)^2-1}}\,\{B_0(u)\,\alpha
_0(u)\,\sin\frac{u}{\lambda}+D_0(v)\,\beta
_0(v)\,\sin\frac{v}{\lambda}\}\ .
\end{equation}
Since $B_0(0)=D_0(0)=1$ we see that this $V$ matches $V$ in
(\ref{24}) and (\ref{25}) on $v=0$ and $u=0$ respectively. Now we
turn our attention to equation (\ref{30}) to determine $M(u, v)$.
Using (\ref{31}) and (\ref{39}) we can write this equation as
\begin{equation}\label{40}
2\,M_{uv}+U_{uv}=\frac{4\,B_0\,D_0\,\alpha _0\,\beta
_0}{(B_0^2+D_0^2-1)}\,\cos\frac{u}{\lambda}\,\cos\frac{v}{\lambda}+O(\lambda
)\ ,
\end{equation}
from which we obtain
\begin{equation}\label{41}
2\,M+U=\frac{4\,\lambda ^2\,B_0\,D_0\,\alpha _0\,\beta
_0}{(B_0^2+D_0^2-1)}\,\sin\frac{u}{\lambda}\,\sin\frac{v}{\lambda}+F_\lambda
(u)+G_\lambda (v)+O(\lambda ^3)\ ,
\end{equation}
where $F_\lambda , G_\lambda$ are functions of integration. By
(\ref{24}) and (\ref{25}) we must have $M$ vanishing when $u=0$
and when $v=0$. Using the exact expression for $U$ in (\ref{37})
we find that
\begin{equation}\label{42}
F_\lambda (u)+G_\lambda (v)=-2\,\log (B_\lambda (u)\,D_\lambda
(v))+O(\lambda ^3)\ .
\end{equation}
Thus $M$ is given by
\begin{equation}\label{43}
M=-\log\frac{B_\lambda\,D_\lambda}{\sqrt{B_\lambda ^2+D_\lambda
^2-1}}+\frac{2\,\lambda ^2\,B_0\,D_0\,\alpha _0\,\beta
_0}{(B_0^2+D_0^2-1)}\,\sin\frac{u}{\lambda}\,\sin\frac{v}{\lambda}+O(\lambda
^3)\ .
\end{equation}
Now using (\ref{12}) and (\ref{37}) we can write this as
\begin{equation}\label{44}
{\rm
e}^{-M}=\frac{B_0(u)\,D_0(v)}{\sqrt{B_0^2(u)+D_0^2(v)-1}}+O(\lambda
^2)\ .
\end{equation}
We note that (\ref{44}) is sufficiently accurate for substitution
into the line--element (\ref{23}) for $u>0,\ v>0$ but (\ref{43})
is needed to verify that the field equation (\ref{30}) is
satisfied. Substitution of (\ref{38}), (\ref{39}) and (\ref{44})
into (\ref{23}) gives the line--element of the space--time in the
region $u>0,\ v>0$ after the collision of the waves as
\begin{equation}\label{45}
ds^2=ds_0^2-(B_0^2(u)+D_0^2(v)-1)\,V\,(dx^2-dy^2)+O(\lambda ^2)\ ,
\end{equation}
with $V=O(\lambda )$ given by (\ref{39}). This is a small
perturbation of a background space--time with line--element
\begin{equation}\label{46}
ds_0^2=-\{B_0^2(u)+D_0^2(v)-1\}\,(dx^2+dy^2)+
\frac{2\,B_0(u)\,D_0(v)}{\sqrt{B_0^2(u)+D_0^2(v)-1}}\,du\,dv\ .
\end{equation}
With $U, V, M$ given by (\ref{37}), (\ref{38}), (\ref{39}) and
(\ref{44}) we must now substitute these functions into the two
remaining field equations (\ref{27}) and (\ref{28}). We find that
since $B_\lambda (u)$ satisfies (\ref{9}) and (\ref{12}) we have
(\ref{27}) now automatically satisfied and (\ref{28}) is
automatically satisfied because $D_\lambda (v)$ satisfies
(\ref{35}) and (\ref{36}).

The Newman--Penrose components $\Psi _A\ (A=0, 1, 2, 3, 4)$ of the
gravitational field of the waves (the Riemann curvature tensor) in
the post--collision region $u>0,\ v>0$ are given by $\Psi _1=\Psi
_3=0$ and
\begin{eqnarray}\label{47}
\Psi _0&=&\frac{\lambda ^{-1}\,\beta
_0(v)\,D_0(v)}{\sqrt{B_0^2(u)+D_0^2(v)-1}}\,\sin\frac{v}{\lambda}+O(\lambda
^0)\ ,\\ \Psi _2&=&O(\lambda ^0)\ ,\\ \Psi _4&=&\frac{\lambda
^{-1}\,\alpha
_0(u)\,B_0(u)}{\sqrt{B_0^2(u)+D_0^2(v)-1}}\,\sin\frac{u}{\lambda}+O(\lambda
^0)\ .
\end{eqnarray}
Thus for small $\lambda$ the field is dominated by two systems of
waves, one described by $\Psi _0$ corresponding to waves with
propagation direction $\partial /\partial u$ in the space--time
and the other described by $\Psi _4$ corresponding to waves with
propagation direction $\partial /\partial v$ in the space--time.
It looks very like a straightforward superposition of waves
traveling in opposite directions were it not for the presence of
the factor $(B_0^2(u)+D_0^2(v)-1)^{-1/2}$. Through this factor
each wave system interferes with the other and a curvature
singularity appears when \begin{equation}\label{50}
B_0^2(u)+D_0^2(v)=1\ .
\end{equation}
This is where the two systems of `rays' (the integral curves of
the vector fields $\partial /\partial u$ and $\partial /\partial
v$) converge. The appearance of a curvature singularity following
the collision is to be expected \cite{G}.

The background space--time corresponding to the post--collision
space--time region $u>0,\ v>0$ has line--element (\ref{46}). This
is not a vacuum space--time. Its Ricci tensor has components
\begin{equation}\label{51}
R^{(0)}_{ij}=-\frac{\alpha
_0^2(u)\,B_0^2(u)}{(B_0^2(u)+D_0^2(v)-1)}\,u_{,i}\,u_{,j}-\frac{\beta
_0^2(v)\,D_0^2(v)}{(B_0^2(u)+D_0^2(v)-1)}\,v_{,i}\,v_{,j}\ .
\end{equation} This would be a simple superposition of the
background Ricci tensors associated with the incoming high
frequency gravitational waves were it not for the factor
$(B_0^2(u)+D_0^2(v)-1)^{-1}$ in each term. The Weyl tensor  of
this background has only one non--vanishing Newman--Penrose
component
\begin{equation}\label{52}
\Psi _2=-\frac{B_0(u)\,D_0(v)\,\dot B_0(u)\,\dot
D_0(v)}{(B_0^2(u)+D_0^2(v)-1)^2}\ .
\end{equation}
This is a type D Weyl tensor with $\partial /\partial u,\ \partial
/\partial v$ as the two degenerate principal null directions. We
see that (\ref{50}) is a curvature singularity in this space--time
also.

 \setcounter{equation}{0}
\section{Collision of Electromagnetic Waves}\indent
We now consider the head--on collision of linearly polarized high
frequency electromagnetic waves. In this case the line--element
(\ref{23}) in the region of space--time $u\geq 0,\ v\leq 0$ is
taken to be the line--element (\ref{17}) with $B(u)=B_\lambda
(u)$, where $\ddot B_\lambda +\phi _2^2B_\lambda =0$ with $\phi
_2(u)$ given by (\ref{20}). Thus in this region
\begin{equation}\label{53}
U=-2\,\log B_\lambda (u)\ ,\ V=M=0\ ,\ \phi
_2=a_0(u)\,\cos\frac{u}{\lambda}\ .\end{equation} For $u\leq 0,\
v\geq 0$ we have
\begin{equation}\label{54}
U=-2\,\log D_\lambda (v)\ ,\ V=M=0\ ,\end{equation}and in this
case the only non--vanishing Newman--Penrose component of the
electromagnetic field is
\begin{equation}\label{55}
\phi _0(v)=b_0(v)\,\cos\frac{v}{\lambda}\ .
\end{equation}
As before we take the region $u\leq 0,\ v\leq 0$ to be flat
space--time with $U=V=M=0$ there. The functions $B_\lambda (u),\
D_\lambda (v)$ for $\lambda\geq 0$ have the properties (\ref{12})
and (\ref{36}) with $B_0(u)$ satisfying (\ref{22a}) and $D_0(v)$
satisfying
\begin{equation}\label{55'}
\ddot D_0+\frac{1}{2}b_0^2D_0=0\ .
\end{equation}
Also (\ref{33}) and (\ref{34}) continue to hold with $\alpha _0,\
\beta _0$ replaced by $a_0,\ b_0$ respectively.

We look for approximate solutions of the Einstein--Maxwell vacuum
field equations in the post--collision region $u>0,\ v>0$. This
involves determining $U, V, M, \phi _0, \phi _2$ for $u>0,\ v>0$
satisfying the boundary conditions: when $u\geq 0,\ v=0$ these
functions are given by (\ref{53}) together with $\phi _0=0$, and
when $u=0,\ v\geq 0$ they are given by (\ref{54}) and (\ref{55})
along with $\phi _2=0$. The field equations to be satisfied are
Einstein's equations with an electromagnetic field as source:
\begin{equation}\label{56}
2\,U_{uu}-U_u^2-V_u^2+2\,U_u\,M_u-4\,\phi _2^2=O(\lambda )\ ,
\end{equation}
\begin{equation}\label{57}
2\,U_{vv}-U_v^2-V_v^2+2\,U_v\,M_v-4\,\phi _0^2=O(\lambda )\ ,
\end{equation}
\begin{equation}\label{58}
2\,V_{uv}-U_u\,V_v-U_v\,V_u-4\,\phi _0\,\phi _2=O(\lambda )\ ,
\end{equation}
\begin{equation}\label{59}
2\,M_{uv}+U_u\,U_v-V_u\,V_v=O(\lambda )\ ,
\end{equation}
and \begin{equation}\label{60} U_{uv}-U_u\,U_v=O(\lambda )\ ,
\end{equation}
together with Maxwell's equations
\begin{equation}\label{61}
\frac{\partial\phi _0}{\partial u}-\frac{1}{2}U_u\,\phi
_0+\frac{1}{2}V_v\,\phi _2=O(\lambda )\ ,
\end{equation}
\begin{equation}\label{62}
\frac{\partial\phi _2}{\partial v}-\frac{1}{2}U_v\,\phi
_2+\frac{1}{2}V_u\,\phi _0=O(\lambda )\ .
\end{equation}
The exact vacuum Einstein--Maxwell field equations are given by
(\ref{56})--(\ref{62}) with the $O(\lambda )$--terms replaced by
zeros \cite{G}. Just as in the gravitational case we conclude
immediately that
\begin{equation}\label{63}
{\rm e}^{-U}=B_\lambda ^2(u)+D_\lambda
^2(v)-1=B_0^2(u)+D_0^2(v)-1+O(\lambda ^2)\ .
\end{equation} Working through the equations
(\ref{56})--(\ref{59}) and (\ref{61}), (\ref{62}) we find that in
$u>0,\ v>0$,
\begin{equation}\label{64}
V=\frac{2\,\lambda
^2B_0(u)\,D_0(v)\,a_0(u)\,b_0(v)}{B_0^2(u)+D_0^2(v)-1}
\,\sin\frac{u}{\lambda}\,\sin\frac{v}{\lambda}\ ,
\end{equation}
\begin{equation}\label{65}
\phi
_0=\frac{D_0(v)\,b_0(v)}{\sqrt{B_0^2(u)+D_0^2(v)-1}}\,\cos\frac{v}{\lambda}
\ ,
\end{equation}
\begin{equation}\label{66}
\phi
_2=\frac{B_0(u)\,a_0(u)}{\sqrt{B_0^2(u)+D_0^2(v)-1}}\,\cos\frac{u}{\lambda}
\ ,
\end{equation}
\begin{equation}\label{67}
{\rm e}^{-M}=\frac{B_\lambda (u)\,D_\lambda (v)}{\sqrt{B_\lambda
^2(u)+D_\lambda ^2(v)-1}}+O(\lambda ^3)\ .
\end{equation} Clearly from (\ref{65}) and (\ref{66}) there are
electromagnetic waves in this region traveling in opposite
directions. They are not a superposition of the incoming
electromagnetic waves however and they are singular on
$B_0^2(u)+D_0^2(v)=1$. The line--element of this post--collision
region of the space--time is given by $ds^2=ds^2_0+O(\lambda ^2)$.
This is in the form of a background having line--element $ds_0^2$
given again by (\ref{46}) and a small \emph{second} order
perturbation. The Weyl tensor of this region of space--time has
non--zero Newman--Penrose components
\begin{equation}\label{69}
\Psi _0=\Psi
_4=\frac{B_0(u)\,D_0(v)\,a_0(u)\,b_0(v)}{B_0^2(u)+D_0^2(v)-1}
\,\sin\frac{u}{\lambda}\,\sin\frac{v}{\lambda}+O(\lambda
)=O(\lambda ^0)\ ,
\end{equation}
\begin{equation}\label{69'}
\Psi _2=-\frac{B_0(u)\,D_0(v)\,\dot B_0(u)\,\dot
D_0(v)}{(B_0^2(u)+D_0^2(v)-1)^2}+O(\lambda )=O(\lambda ^0)\
.\end{equation}Thus the collision of the electromagnetic waves has
produced two systems of identical gravitational waves traveling in
opposite directions together with a comparable magnitude `Coulomb'
term (\ref{69'}).

We can easily combine the results of this section with those of
section 3. If the incoming high frequency, linearly polarized
plane gravitational waves share their wave fronts with the high
frequency, linearly polarized plane electromagnetic waves then the
line--element in the region $u\geq 0,\ v\leq 0$ is given by
(\ref{7}) and (\ref{8}) with $\phi _2$ given by (\ref{20}). Now
(\ref{9}) is replaced by
\begin{equation}\label{70}
\ddot B_\lambda +(\dot w_\lambda ^2+\phi _2^2)\,B_\lambda =0\ .
\end{equation}
As before we have $B_\lambda (u)=B_0(u)+O(\lambda ^2)$ where now
$B_0(u)$ satisfies
\begin{equation}\label{71}
\ddot B_0+\frac{1}{2}(\alpha _0^2+a_0^2)\,B_0=0\ .
\end{equation}
In this case we have in the region $u\geq 0,\ v\leq 0$
\begin{equation}\label{72}
U=-2\,\log B_\lambda (u)\ ,\ V=2\,\lambda\,\alpha
_0(u)\,\sin\frac{u}{\lambda}\ ,\ M=0\ ,\ \phi
_2=a_0(u)\,\cos\frac{u}{\lambda}\ .
\end{equation} In the region $u\leq 0,\ v\geq 0$ we have
\begin{equation}\label{73}
U=-2\,\log D_\lambda (v)\ ,\ V=2\,\lambda\,\beta
_0(v)\,\sin\frac{v}{\lambda}\ ,\ M=0\ ,\ \phi
_0=b_0(v)\,\cos\frac{v}{\lambda}\ ,
\end{equation} with $D_\lambda (v)=D_0(v)+O(\lambda ^2)$ and
$D_0(v)$ satisfying
\begin{equation}\label{74}
\ddot D_0+\frac{1}{2}(\beta _0^2+b_0^2)\,D_0=0\ .
\end{equation}
Now the space--time in the post collision region has line--element
(\ref{23}) with $U$ given by (\ref{38}) and
\begin{eqnarray}\label{75}
V&=&\frac{2\,\lambda}{\sqrt{B_0^2(u)+D_0^2(v)-1}}\,\{B_0(u)\,\alpha
_0(u)\,\sin\frac{u}{\lambda}+D_0(v)\,\beta
_0(v)\,\sin\frac{v}{\lambda}\}\nonumber\\ &&+\frac{2\,\lambda
^2B_0(u)\,D_0(v)\,a_0(u)\,b_0(v)}{B_0^2(u)+D_0^2(v)-1}\,\sin\frac{u}{\lambda}\,
\sin\frac{v}{\lambda}\ ,
\end{eqnarray}
\begin{eqnarray}\label{76}
\phi
_0&=&\frac{D_0(v)\,b_0(v)}{\sqrt{B_0^2(u)+D_0^2(v)-1}}\,\cos\frac{v}{\lambda}\nonumber\\
&&-\frac{\lambda\,D_0(v)\,B_0(u)\,\beta
_0(v)\,a_0(u)}{B_0^2(u)+D_0^2(v)-1}
\,\cos\frac{v}{\lambda}\,\sin\frac{u}{\lambda}+O(\lambda ^2)\ ,
\end{eqnarray}
\begin{eqnarray}\label{77}
\phi
_2&=&\frac{B_0(u)\,a_0(u)}{\sqrt{B_0^2(u)+D_0^2(v)-1}}\,\cos\frac{u}{\lambda}\nonumber\\
&&-\frac{\lambda\,D_0(v)\,B_0(u)\,\alpha
_0(u)\,b_0(v)}{B_0^2(u)+D_0^2(v)-1}
\,\cos\frac{u}{\lambda}\,\sin\frac{v}{\lambda}+O(\lambda ^2)\ ,
\end{eqnarray}
and $M$ is given by (\ref{67}). This approximate solution of the
Einstein--Maxwell equations gives the post collision region of the
space--time in such a form that all possible combinations of pairs
of colliding systems of waves can be extracted as special cases.
For example, the collision of high frequency, linearly polarized
plane electromagnetic waves with high frequency, linearly
polarized plane gravitational waves corresponds to putting $\alpha
_0=b_0=0$; after collision these would superimpose were it not for
the factor $(B_0^2(u)+D_0^2(v)-1)^{-1/2}$ appearing in $V$ and
$\phi _2$.

\setcounter{equation}{0}
\section{Arbitrary Polarization}\indent
All of the foregoing can be generalized to allow the gravitational
and electromagnetic waves to have the maximum two degrees of
freedom of polarization. We have already pointed out following
(\ref{18}) how to introduce this generalization in the case of
electromagnetic waves. If the plane gravitational waves have
general polarization then the line--element (\ref{3}) must be
generalized to the Rosen--Szekeres form
\begin{eqnarray}\label{78}
ds^2&=&-B^2\{({\rm e}^w\,\cosh q\,dx+{\rm e}^{-w}\,\sinh
q\,dy)^2+({\rm e}^w\,\sinh q\,dx+{\rm e}^{-w}\,\cosh
q\,dy)^2\}\nonumber\\&& +2\,du\,dv\ ,
\end{eqnarray}
with $B, w, q$ functions of $u$ only. How this arises is outlined
in some detail in \cite{FH}. The Ricci tensor calculated with this
metric tensor has components
\begin{equation}\label{79}
R_{ij}=2\,\left (\frac{\ddot B}{B}+\dot w^2\cosh ^22\,q+\dot
q^2\right )\,u_{,i}\,u_{,j}\ \ ,
\end{equation} and the only non--vanishing Weyl tensor component
in Newman--Penrose notation is
\begin{eqnarray}\label{80}
\Psi _4&=&-\ddot w\,\cosh 2\,q-4\,\dot w\,\dot q\,\sinh
2\,q-2\,B^{-1}\dot B\,\dot w\,\cosh 2\,q\nonumber\\ && +i\,(\ddot
q+2\,B^{-1}\dot B\,\dot q-\dot w^2\sinh 4\,q)\ .
\end{eqnarray} Following the Burnett \cite{B} approach described
in section 2 we specialize to high frequency plane waves by making
the replacements $B(u)\rightarrow B_\lambda (u),\ w(u)\rightarrow
w_\lambda (u),\ q(u)\rightarrow q_\lambda (u)$ with these
functions satisfying the Einstein vacuum field equation
\begin{equation}\label{81}
\ddot B_\lambda +(\dot w_\lambda ^2\cosh ^22\,q_\lambda +\dot
q_\lambda ^2)\,B_\lambda =0\ ,
\end{equation}
and \begin{equation}\label{82} w_\lambda (u)+i\,q_\lambda
(u)=\lambda\,\alpha _0(u)\,\sin\frac{u}{\lambda}\ ,
\end{equation}
where now $\alpha _0(u)$ is a \emph{complex}--valued function of
the real variable $u$. Following the procedure outlined after
equation (\ref{9}) [see reference \cite{FH}] we have $B_\lambda
(u)=B_0(u)+O(\lambda ^2)$ and $B_0(u)$ satisfies
\begin{equation}\label{83}
\ddot B_0+\frac{1}{2}|\alpha _0|^2B_0=0\ .
\end{equation}
The line--element (\ref{13}) is now replaced by
\begin{equation}\label{84}
ds^2=ds_0^2-2\,\lambda\,B_0^2(u)\,\sin\frac{u}{\lambda}\,\Re\{\alpha
_0(u)\,(dx-idy)^2\}+O(\lambda ^2)\ ,
\end{equation} where $\Re$ denotes the real part of the quantity
following it in brackets and $ds_0^2$ is given by (\ref{14}). The
background Ricci tensor (\ref{15}) reads now
\begin{equation}\label{85}
R^{(0)}_{ij}=-|\alpha _0(u)|^2u_{,i}\,u_{,j}\ \ ,
\end{equation}
while the gravitational field (\ref{16}) is generalized to
\begin{equation}\label{86}
\Psi _4=\frac{1}{\lambda}\,\bar\alpha
_0(u)\,\sin\frac{u}{\lambda}+O(\lambda ^0)\ ,
\end{equation}
with the bar denoting complex conjugation. The fact that $\Psi _4$
here is complex shows that we are dealing with waves having two
degrees of freedom of polarization.

If we wish to include high frequency plane electromagnetic waves
of arbitrary polarization sharing their wave fronts with these
gravitational waves then such waves have a Maxwell field with one
non--vanishing Newman--Penrose component
\begin{equation}\label{87}
\phi _2(u)=a_0(u)\,\cos\frac{u}{\lambda}\ ,
\end{equation}
where now $a_0(u)$ is a \emph{complex}--valued function of the
real variable $u$. The equations (\ref{84})--(\ref{86}) [the
latter is now the Weyl conformal curvature tensor] continue to
hold but (\ref{81}) is replaced by
\begin{equation}\label{87'}\ddot B_\lambda +(\dot w_\lambda
^2\cosh ^22\,q_\lambda +\dot q_\lambda ^2+|\phi _2|^2)\,B_\lambda
=0\ ,
\end{equation}and (\ref{83}) is replaced by
\begin{equation}\label{88}
\ddot B_0+\frac{1}{2}(|\alpha _0|^2+|a_0|^2)\,B_0=0\ ,
\end{equation}
and so (\ref{85}) becomes in this case
\begin{equation}\label{89}
R^{(0)}_{ij}=-(|\alpha _0|^2+|a_0|^2)\,u_{,i}\,u_{,j}\ \ .
\end{equation}

The line--element in the post collision part of the space--time in
this case is the generalized Rosen--Szekeres form
\begin{eqnarray}\label{90}
ds^2&=&-{\rm e}^{-U}\{{\rm e}^V\,\cosh W\,dx^2-2\,\sinh
W\,dx\,dy+{\rm e}^{-V}\,\cosh W\,dy^2\}\nonumber\\ &&+2\,{\rm
e}^{-M}\,du\,dv\ ,
\end{eqnarray}where $U, V, W, M$ are functions of $u, v$. This
coincides with (\ref{23}) when $W=0$. As initial data on $u\geq
0,\ v=0$ we have
\begin{equation}\label{91}
U=-2\,\log B_\lambda (u)\ ,\ V-iW=2\,\lambda\,\alpha
_0(u)\,\sin\frac{u}{\lambda}\ ,\ M=0\ ,
\end{equation}
and
\begin{equation}\label{91'}
\phi _2=a_0(u)\,\cos\frac{u}{\lambda}\ ,\ \phi _0=0\
.\end{equation} The data on $u=0,\ v\geq 0$ is
\begin{equation}\label{92}
U=-2\,\log D_\lambda (v)\ ,\ V-iW=2\,\lambda\,\beta
_0(v)\,\sin\frac{v}{\lambda}\ ,\ M=0\ ,
\end{equation}
and
\begin{equation}\label{92'}
\phi _0=b_0(v)\,\cos\frac{v}{\lambda}\ ,\ \phi _2=0\
.\end{equation}Here $\beta _0,\ b_0$ are complex--valued functions
and $D_\lambda (v)=D_0(v)+O(\lambda ^2)$ with $D_0(v)$ satisfying
\begin{equation}\label{93}
\ddot D_0+\frac{1}{2}(|\beta _0|^2+|b_0|^2)\,D_0=0\ .
\end{equation}We note that $B_\lambda (0)=D_\lambda (0)=1$ and
$\dot B_\lambda (0)=\dot D_\lambda (0)=0$ for $\lambda\geq 0$
while $\alpha _0(0)=\beta _0(0)=a_0(0)=b_0(0)=0$.

Solving this initial value problem for the functions $U, V, W, M$
of $(u, v)$ in the line--element (\ref{90}), and for the Maxwell
field described by $\phi _0(u, v)$ and $\phi _2(u, v)$, in the
region $u>0,\ v>0$, we find that $U$ is once again given by
(\ref{63}) while
\begin{eqnarray}\label{94}
V-iW&=&\frac{2\,\lambda}{\sqrt{B_0^2(u)+D_0^2(v)-1}}\,\{D_0(v)\,\beta
_0(v)\,\sin\frac{v}{\lambda}+B_0(u)\,\alpha
_0(u)\,\sin\frac{u}{\lambda}\}\nonumber\\ &&+\frac{2\,\lambda
^2\,B_0(u)\,D_0(v)\,\bar a_0(u)\,b_0(v)}{B_0^2(u)+D_0^2(v)-1}\,
\sin\frac{u}{\lambda}\,\sin\frac{v}{\lambda}\ ,
\end{eqnarray}
\begin{eqnarray}\label{95}
M&=&-\log\frac{B_\lambda (u)\,D_\lambda (v)}{\sqrt{B_\lambda
^2(u)+D_\lambda ^2(v)-1}}\nonumber\\ &&+\frac{\lambda
^2B_0(u)\,D_0(v)\,(\alpha _0(u)\,\bar\beta _0(v)+\bar\alpha
_0(u)\,\beta
_0(v))}{B_0^2(u)+D_0^2(v)-1}\,\sin\frac{u}{\lambda}\,\sin\frac{v}{\lambda}\nonumber\\
&&+ O(\lambda ^3)\ ,
\end{eqnarray}
\begin{equation}\label{96}
\phi
_0=\frac{b_0(v)\,D_0(v)}{\sqrt{B_0^2(u)+D_0^2(v)-1}}\,\cos\frac{v}{\lambda}
-\frac{\lambda\,B_0(u)\,D_0(v)\,\beta
_0(v)\,a_0(u)}{B_0^2(u)+D_0^2(v)-1}\,\cos\frac{v}{\lambda}\,\sin\frac{u}{\lambda}\
,
\end{equation}
\begin{equation}\label{97}
\phi
_2=\frac{a_0(u)\,B_0(u)}{\sqrt{B_0^2(u)+D_0^2(v)-1}}\,\cos\frac{u}{\lambda}
-\frac{\lambda\,B_0(u)\,D_0(v)\,\bar\alpha
_0(u)\,b_0(v)}{B_0^2(u)+D_0^2(v)-1}\,\cos\frac{u}{\lambda}\,\sin\frac{v}{\lambda}\
.
\end{equation}
These functions satisfy the initial data given above and satisfy
approximately the vacuum Einstein--Maxwell field equations
\cite{G}. Since $Z\equiv V-iW=O(\lambda )$ the field equations
simplify in this case to
\begin{equation}\label{98}
2\,Z_{uv}-U_u\,Z_v-U_v\,Z_u-4\,\phi _0\,\bar\phi _2=O(\lambda )\ ,
\end{equation}
\begin{equation}\label{99}
2\,U_{uu}-U_u\,(U_u-2\,M_u)-|Z_u|^2-4\,|\phi _2|^2=O(\lambda )\ ,
\end{equation}
\begin{equation}\label{100}
2\,U_{vv}-U_v\,(U_v-2\,M_v)-|Z_v|^2-4\,|\phi _0|^2=O(\lambda )\ ,
\end{equation}
\begin{equation}\label{101}
2\,M_{uv}+U_u\,U_v-\Re\{Z_u\,\bar Z_v\}=O(\lambda )\ ,
\end{equation}
together with (\ref{31}) and Maxwell's equations, which now read
\begin{equation}\label{102}
\frac{\partial\phi _0}{\partial u}-\frac{1}{2}U_u\,\phi
_0+\frac{1}{2}Z_v\,\phi _2=O(\lambda )\ ,
\end{equation}
\begin{equation}\label{103}
\frac{\partial\phi _2}{\partial v}-\frac{1}{2}U_v\,\phi
_2+\frac{1}{2}\bar Z_u\,\phi _0=O(\lambda )\ .
\end{equation}The line--element (\ref{90}) becomes
\begin{equation}\label{104}
ds^2=ds_0^2-(B_0^2(u)+D_0^2(v)-1)\,\Re\{Z\,(dx-idy)^2\}+O(\lambda
^2)\ ,
\end{equation}
with $ds_0^2$ given by (\ref{46}). The Weyl tensor is a
generalization of (3.25)--(3.27) with the dominant Newman--Penrose
components given by
\begin{eqnarray}\label{105}
\Psi _0&=&\frac{\lambda ^{-1}\,\beta
_0(v)\,D_0(v)}{\sqrt{B_0^2(u)+D_0^2(v)-1}}\,\sin\frac{v}{\lambda}+O(\lambda
^0)\ ,\\\Psi _4&=&\frac{\lambda ^{-1}\,\bar\alpha
_0(u)\,B_0(u)}{\sqrt{B_0^2(u)+D_0^2(v)-1}}\,\sin\frac{u}{\lambda}+O(\lambda
^0)\ .
\end{eqnarray}

\noindent
\section*{Acknowledgment}\noindent
One of us (D.M.W.) wishes to thank IRCSET for financial support.

\end{document}